\newcounter{myctr}

\documentclass{ws-acs}
\usepackage{subfigure}
\usepackage{cite}

\begin{document}

\markboth{Thimo Rohlf and Chris Winkler}{Emergent network structure 
and evolvable robustness}

\title{EMERGENT NETWORK STRUCTURE, EVOLVABLE ROBUSTNESS AND NON-LINEAR
EFFECTS OF POINT MUTATIONS IN AN ARTIFICIAL GENOME MODEL}

\author{\footnotesize THIMO ROHLF}

\address{Epigenomics program, Genopole, 523 Terasses de l'Agora, F-91000 Evry, France\\
Max-Planck-Institute for Mathematics in the Sciences, \\Inselstr. 22, D-04103 Leipzig, Germany
\\
rohlf@santafe.edu}

\author{CHRIS WINKLER}

\address{Pioneer Hi-Bred International, 7250 NW 62nd Ave., \\Johnston, IA 50131, USA\\
chris.winkler@pioneer.com}

\maketitle

\begin{abstract}
Genetic regulation is a key component in development, but a clear
understanding of the structure and dynamics of genetic networks is not
yet at hand. In this paper we investigate these properties within an
artificial genome model originally introduced by Reil \cite{Reil1999}.
We analyze statistical properties of randomly generated genomes
both on the sequence- and network level, and 
show that this model correctly predicts the
frequency of genes in genomes as found in experimental data. 
Using an evolutionary algorithm based on stabilizing selection
for a phenotype, we show that dynamical robustness against single base mutations,
as well as against random changes in initial states 
of regulatory dynamics that mimic stochastic fluctuations in 
environmental conditions, can emerge in parallel. 
Point mutations at the sequence level have strongly non-linear
effects on network wiring, including as well structurally neutral mutations
and simultaneous rewiring of multiple connections,
which occasionally lead to strong reorganization
of the attractor landscape and metastability of evolutionary dynamics.
Evolved genomes exhibit
characteristic patterns on both sequence and network level.
\end{abstract}


\section{Introduction}
The transcription of DNA into mRNA and subsequent translation into
protein is the fundamental genetic process; it is the crucial first
step by which genetic information gives rise to an
organism. Development is not such a linear process, however. By
binding to specific regions of the genome, the protein produced by one
gene can affect the activity of other genes, and those genes may in
turn express proteins that enhance or inhibit still more genes. A
network of interactions responsible for the regulation of genetic
activity is thus defined. Such genetic regulation is important if
cells are to have independent control over their behavior.

Today, the available amount of data for regulatory interactions
in a number of model organisms, as, for example, Yeast \cite{Wagner2000}
is steadily increasing. A number of distinguishing
structural properties have been identified, namely scale-free
degree distributions \cite{Jordan2004}, motifs \cite{Dobrin2004} and modular organization \cite{Thieffry1999}.

Still, there is not enough information
to suggest a comprehensive theory of how genetic regulatory networks
attain a particular structure, how genes in such networks interact and
respond to perturbation, and how evolution has shaped these
factors. This study is an attempt to explore these questions in the
context of one particular model \cite{Reil1999}, in the hopes that it has
features that correspond to the limited data currently available, and
so that some progress toward a comprehensive theory might eventually
be made.

Traditionally, attempts to understand the characteristics of
regulatory networks have focused on dynamical properties. That is, a
network topology is specified and rules are applied to describe how
each gene in the network responds to inputs. Some initial state is
then assigned and the time evolution of gene activity is studied. A
variety of rules have been used, including Boolean switches
\cite{Kauffman1969}, thresholds \cite{Kurten1988b,Rohlf2002}, and differential equations
\cite{Glass1973}. Much less work has been done in understanding how the
machinery of transcription, translation, and binding might act
throughout the genome to produce the topology of a genetic network. In
fact, most studies of genetic networks ignore modeling DNA-specific
processes altogether \cite{Jong2002}. The first part of our study
examines to what extent Reil's model \cite{Reil1999}, which includes
explicit parameterizations for transcription and translation, can
produce realistic genetic networks based on random genome realizations.

A description of the method we will use for building genetic
regulatory networks follows, along with comparisons to published and
publically available experimental data. 
Statistical properties of random realizations of artificial genomes
are derived, and related to network structure.
Next, we investigate the
dynamics of our modeled networks when applying threshold dynamics to
gene behavior. Although this is a strong simplification, this type of
discrete dynamics has been successfully applied in a  number of studies that
are concerned with the co-evolution of network dynamics and -structure
\cite{Bornholdt1998,Bornholdt2000a,Ciliberti2007}. Finally, we are
interested in understanding the role evolution might play in selecting
particular network topologies. 
This is explored by asking how genome
structure changes when those networks with certain dynamical
properties are preferentially selected. 
Similar questions have been addressed in a small number of
previous proof-of-principle studies using artificial genomes\cite{Banzhaf2003a,Kuo2004a,Hallinan2004a}, however,
without relating the observed adaptation to changes in sequence
and network topology.
In particular, we investigate a
scenario of stabilizing selection similar to previous studies concerned
with the evolution of developmental canalization \cite{Ciliberti2007}, and
evolution of gene regulatory networks in changing environments \cite{}.

 We find evolution
towards robustness of regulatory dynamics against both noise, modeled as fluctuating initial conditions,
and against mutations. We show that, in principle, this phenotypic robustness
can be traced back to adaptive changes on the sequence level
that lead to emergence of more robust regulatory networks. 

\begin{figure}[th]
\centerline{\psfig{file=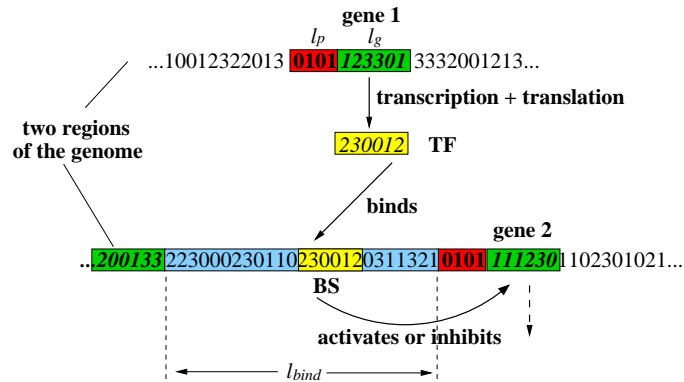,width=9cm}}
\vspace*{8pt}
\caption{\label{fig:schematic} Schematic description of artificial genome construction (after \cite{Reil1999}):
a fixed sequence of $l_p$ digits - here '0101' (red boxes), $l_p =4$ -  is defined as promoter sequence.
Wherever it occurs on the string defining the genome, the next $l_g$ digits
are defined as "genes" (here, $l_g =6$, green boxes). If the gene is active, a transcription
factor (TF) is produced by increasing each digit of the gene sequence by $1$ (yellow), crudely
mimicking transcription and translation. The algorithm searches matching sequences anywhere
in the genome in binding regions (blue box in the lower string) between genes, defining binding sites (BS) for the TF. The so-defined directed regulatory
interaction from gene 1 to gene 2 can either activate or inhibit transcription of the next gene
downstream (gene 2, dashed arrow). For details, see section \ref{ag_construction}
in the text.}
\end{figure}

\section{Model Details}

\subsection{Regulatory network construction from random sequences}\label{ag_construction}
An artificial genome can be constructed as follows (also see
Fig.~\ref{fig:schematic}). Randomly string together $S$ integers drawn
uniformly between 0 and 3. The use of 4 digits need not be the case,
but does provide correspondence with the ATGC alphabet of real
genomes. 
For the purpose of generalization, the length of the alphabet
in the artificial genome may in principle take any positive integer value $\lambda$.
Next, define a base promoter sequence of length $l_p$ to
indicate the position of genes in the genome, say '0101'. Wherever
the promoter sequence occurs, the next $l_{g}$ digits are specified as
a gene. Translation of the gene sequence into a protein occurs
simply. Each number in the sequence is incremented by 1 and any values
greater than the last number in the base set of digits become the
first number ({\it e.g.}, the gene '012323' becomes the protein
'123030'). Binding sites are determined by searching the genome for
the protein sequence. If a match is found, then the protein is a
transcription factor (TF) that binds to that site and that regulates
the next downstream gene. In case there are multiple binding sites
of this TF for this gene, only one of them is counted for
network construction.
TFs may enhance or inhibit gene activity. In
this study each TF has {\it equal} contribution to a gene's state and
has {\it equal} probability of activating or suppressing gene
expression. In real genetic systems, a TF may activate some genes
and inhibit others, depending on a complex interplay between various
factors that do not only depend on sequence. In our study,
we make the simplifying assumption that a TF is either activating
or inhibiting, which is determined by the sum $s_g$ of its sequence:
if $s_g < (1/2)s_{max}$, where $s_{max} = (\lambda-1)l_g$ is the maximal possible
cross sum value, it is inhibiting, otherwise it is activating.
Activation and inhibition are reflected by different weight values
in the interaction matrix (network) defined by all TFs and their
corresponding binding sites (cf. section \ref{dyndef}).

Clearly this model greatly simplifies the true transcription,
translation, and binding processes. The binding of a real transcription
factor to a cis-site, for example, depends on the protein's structure,
shape, and environment, rather than a simple template matching
approach. Moreover, there is a stochastic element to all these
processes that is simply ignored here.

Although it represents a strong simplification, the model does have biological
justification \cite{Reil1999}. The use of a base promoter sequence is
reminiscent of the TATA box frequently found in eukaryotic
organisms. Binding is modeled in a DNA-specific way, just as in real
organisms. Additionally, the model has the potential for greater
extendability than some models ({\it e.g.}, Boolean networks) because
it includes DNA-specific transcription, translation, and binding. The
impact of single base pair mutations on gene function and network
structure can be studied with this model, and also the effect
of sequence duplications (resulting in gene duplication) or -deletions
\cite{Leier2007}. In this paper, we will restrict ourselves
to single base pair mutations, and keep the genome size constant,
both with respect to the number of bases $S$ and the number of genes
$N$.

\subsection{Regulatory dynamics}\label{dyndef}
Dynamics of state changes (activity or inhibition of genes)
on the constructed networks can be defined in various ways. In our
study, we apply random threshold network (RTN) dynamics:
An RTN consists 
of $N$ randomly interconnected binary sites (spins) with states $\sigma_i=\pm1$.
For each site $i$, its state at time $t+1$ is a function of the inputs it receives 
from other spins at time $t$:
\begin{eqnarray} 
\sigma_i(t+1) = \left \{
\begin{array}{ll} +1, & \, f_i(t) > 0\\
                  -1, & \, f_i(t) \le 0
  \end{array}\right.                
\end{eqnarray}  
with 
\begin{eqnarray} 
f_i(t) = \sum_{j=1}^N c_{ij}\sigma_j(t) + h.  
\end{eqnarray}
The $N$ network sites are updated synchronously. In the following
discussion the threshold parameter $h$ is set to zero. The interaction weights
$c_{ij}$ take discrete values $c_{ij} = +1$ (activation) or $-1$ (inhibition);
whether a given interaction is activating or inhibiting, is defined by the TF
it is derived from, as explained in section \ref{ag_construction}. 
If $i$ does not receive signals from $j$, one has $c_{ij} = 0$.

For a finite system size
$N$, the dynamics of RTN, which are closely related to Boolean
networks \cite{Kauffman1969} converge to periodic attractors (limit cycles)
after a finite number of updates. It has been suggested that different limit cycles
may correspond to different gene expression states (cell types) \cite{Kauffman1969}.
This property of RTN is also advantageous for defining phenotypes in
artificial evolutionary scenarios that are subject to various kinds of selective
pressure \cite{Ciliberti2007}.

\section{Statistical properties of the artificial genome}

In the following, $N$ denotes the number of genes in the artificial genome,
$S$ the number of bases, $l_g$ the length of gene sequences, $l_p$ the length of promoter sequences 
(both are fixed and identical for
all genes), and $\lambda$ the length of the alphabet. We now show how these quantities are interrelated 
via the combinatorial
construction of the artificial genome, as outlined in section \ref{ag_construction}.

\subsection{Statistical distribution of $l_{bind}$}
Let us first derive the statistical distribution of lengths $l_{bind}$ of the binding regions
preceding promoters in the artificial genome. We incrementally draw a sequence of random digits (bases) from
the alphabet. 
Once we have drawn at least $s \ge l_p$ bases, 
the probability that a promoter sequence
is generated by chance with base $s$ is $p_{p} = (1/\lambda)^{l_p}$, since the last $l_p$ digits
must have position-specific values according to the predefined promoter sequence, and each of
these values has probability $1/\lambda$ to occur.
Hence, the probability distribution of the number $X$ of Bernoulli trials (i.e.
the sequence length) needed to
get one success (a promoter sequence)
is a geometric distribution for $s \ge l_p$ and zero otherwise,
\begin{equation} \label{s_dist}
P(X=s) = 
\begin{cases}
0 & \text{if $0 < s < l_p$}\\
p_{p}(1-p_{p})^{s-l_p} &  \text{if $s \ge l_p$}
\end{cases}
\end{equation}

Since the last $l_p$ digits constitute the promoter, the length of the preceding binding region
is given by $l_{bind} = s - l_p$, and it follows
\begin{eqnarray} p(l_{bind}) & = & p_p(1-p_p)^{l_{bind}} \nonumber\\
                             & = & \lambda^{-l_p}(1-\lambda^{-l_p})^{l_{bind}} \label{lb_dist}\\
                             & = &  \lambda^{-l_p}\exp{[-\alpha \cdot l_{bind}]},\label{lbinddist_eq}\end{eqnarray}
which is a decaying exponential distribution with $\alpha = -\ln{(1-\lambda^{-l_p})}$.

From Eq. \ref{lb_dist} follows that the average length of binding regions is given by
\begin{equation}  \langle l_{bind} \rangle = \lambda^{l_p} - 1,   \end{equation}
which is the mean of the geometric distribution.

\subsection{Genome size scaling}
From Eq. \ref{s_dist} follows that
on average, we have to draw $\lambda^{l_p} + l_p - 1$ bases to obtain a promoter sequence; the next $l_g$ bases
are defined as the associated gene. Hence, to produce genomes with exactly $N$
genes, the expectation value for the number of bases $S$ that we have to string together
is
\begin{equation} \langle S\rangle = N \cdot(\lambda^{l_p}+ l_p - 1 +l_g).\end{equation}
If we instead keep $S$ fixed and ask for the expected number of genes,
under the assumption that $l_p < l_g \ll \lambda^{l_p}$, which holds for typical values 
considered in this study (e.g., $l_g = 6$, $l_p =4$ and $\lambda = 4$), we conclude
that
\begin{equation} \langle N \rangle \approx \frac{1}{\lambda^{l_p}}\cdot S.\end{equation}

\subsection{Network connectivity}
In this section, we relate the previously derived statistical properties of the artificial genome to characteristic parameters of the resulting random networks.

\subsubsection{Average connectivity}\label{avconn}
For a given TF, the probability to match to a random base sequence of length $l_g$ is given by $p_{bind} = \lambda^{-l_g}$.
There are $n := \langle l_{bind}\rangle-l_g+1$ subsequences of length $l_g$ in a binding region of expected length 
$\langle l_{bind}\rangle$. The probability that none of these matches the TF sequence is
\begin{equation}
p_{0} = (1-p_{bind})^n,
\end{equation}
thus the probability that
the TF provides at least one input 
to the gene defined by the promoter sequence following a binding
region is \footnote{There is a finite chance that the same TF can bind more than once in a given
binding region, however, since the update scheme for network dynamics requires
uniquely defined connections, we assign only one regulatory input in this case. If the distance
between promoters is smaller than $l_g$, no binding occurs, however, for typical parameter values
of $\lambda$, $l_g$ and $l_p$ as applied in our study, this is a very unlikely event and can be neglected.} 
\begin{eqnarray} p_{input} = 1 - p_0 = 1 - (1-\lambda^{-l_g})^{\langle l_{bind}\rangle-l_g+1}.\end{eqnarray}
Since, in a genome with $N$ genes, we have $N$ binding regions and $N$ transcription factors, the total number of regulatory interactions $\langle k_{total}\rangle$ per
genome
(averaged over the whole ensemble of possible random genomes) scales quadratically with the number $N$ of genes,
\begin{equation} \langle k_{total}\rangle = p_{input}\cdot N^2,  \end{equation}
and the slope depends on $\lambda$, $l_g$ and $l_p$. It follows that the average connectivity (wiring density)
$\langle k \rangle := \langle k_{total}\rangle/N$ scales linear with $N$.

Notice, however, that the average  number of regulatory interactions $\bar{K}$ obtained from a {\em particular} genome realization 
can substantially deviate from $\langle k \rangle$, since the possible values of $\bar{K}$ are approximately Gaussian 
distributed \cite{Hallinan2004a}.

\begin{figure}
\centering
\subfigure[outdegree-distribution]{
\resizebox{60mm}{!}{\includegraphics{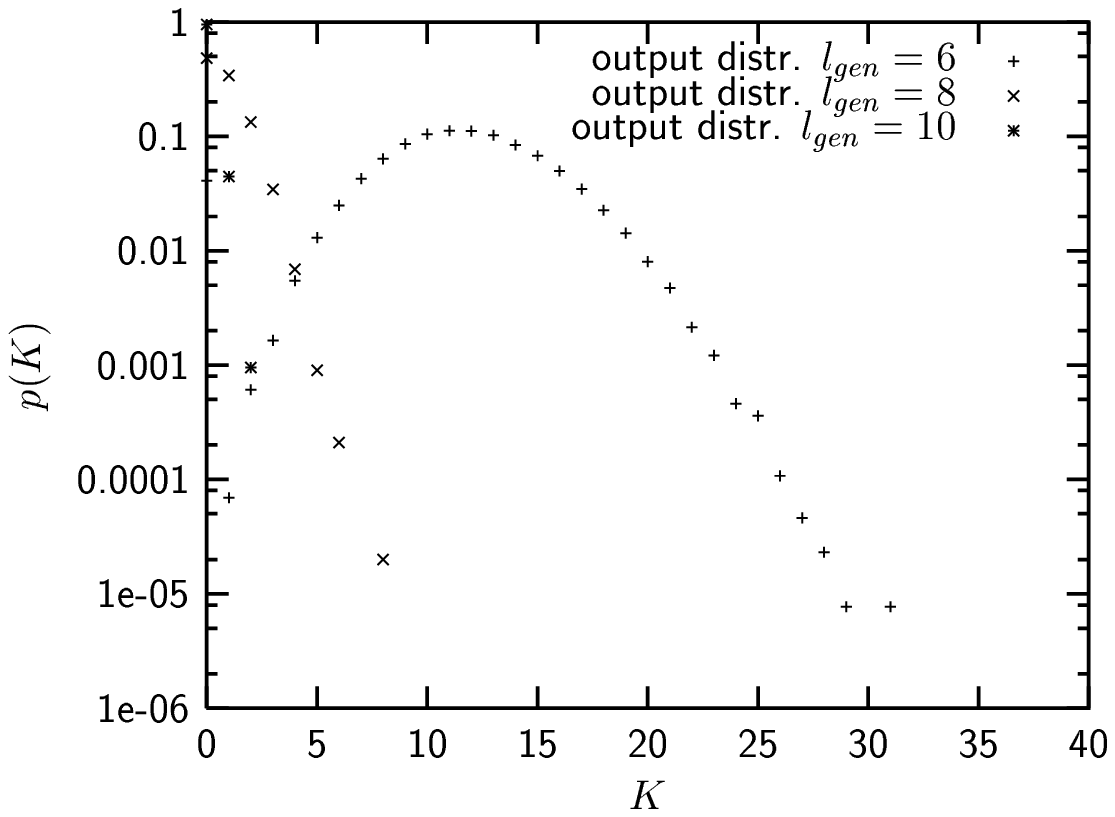}}
}
\subfigure[indegree
-distribution]{
\resizebox{60mm}{!}{\includegraphics{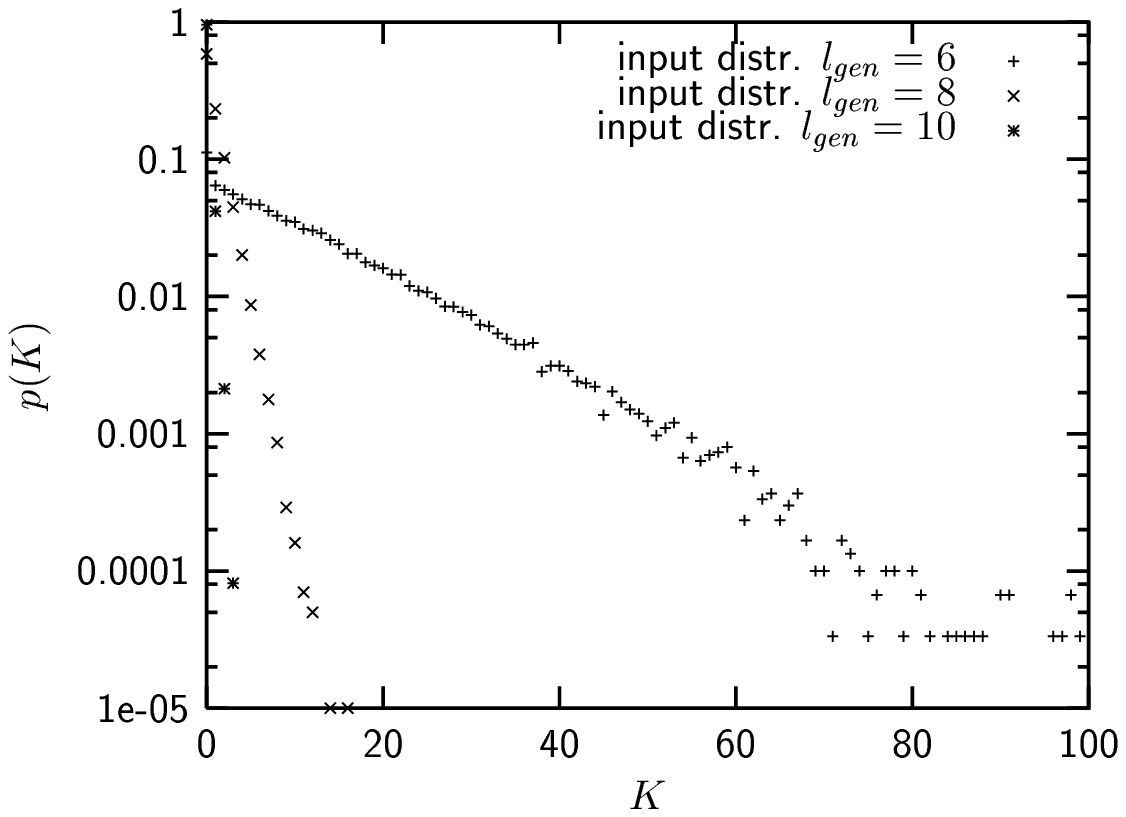}}
}
\caption{The probability of having $K_{out}$ regulatory outputs (a) and the probability of having $K_{in}$ regulatory inputs
for random genomes with different gene lengths $l_g$, averaged over $10^4$ realizations.}
\vspace*{8pt}
\end{figure}

\subsubsection{In- and outdegree distribution}
From the above considerations, it is straight-forward to derive the statistical distributions for the number
of ingoing and outgoing links in randomly constructed genomes. As denoted in section  \ref{ag_construction} (see also Fig. 1),
the subsequent processes of transcription/translation of gene sequences (incrementation of each number in the gene sequence by 1),
defining transcription factors TF, and binding of the TF to subsequences of the base string by template matching,
defines a network of directed regulatory interactions. A given TF represents an out-link of the gene that codes for it,
and an in-link for all other genes that have binding sites for this TF.
By definition of template matching,
each TF has equal probability $p_{bind} = \lambda^{-l_g}$ to bind at any region of the base string
(cf. section \ref{avconn}), and hence generation of out-links is
a Poisson process \cite{ErdHos1960}. Consequently, the outdegree distribution is a Poissonian (Fig. 2a):
\begin{equation}
P(k_{out}) = \frac{\langle K\rangle ^{k_{out}}}{k_{out}!}\exp{[-\langle K\rangle]}.
\end{equation}
The number of inputs a gene receives from other genes, however, is proportional to the length $l_{bind}$ of its associated binding region,
hence, it follows from Eq. \ref{lbinddist_eq}
\begin{equation}
P(k_{in}) \sim \exp{[-\beta k_{in}]} ,
\end{equation}
i.e. the indegree distribution is exponential. 
Both results are confirmed by numerical simulations (Fig. 2a  and 2b).

\begin{figure}[th]
\centerline{\psfig{file=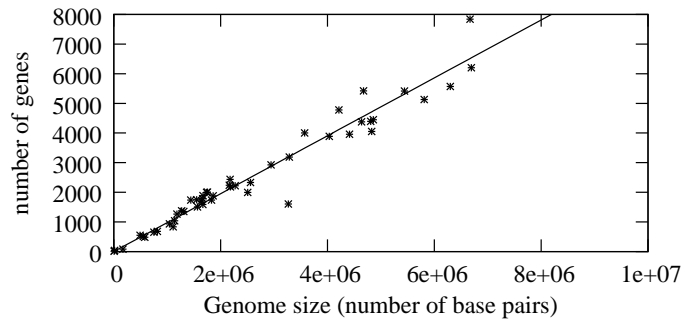,width=9cm}}
\vspace*{8pt}
\caption{\label{fig:numgenes}The number of genes predicted from the model as a function of
genome size $S$ with $l_{p}=5$ (line). Data points (+) show the number of genes in 50 prokaryotic organisms, for which
complete sequence information is available. Observed data are taken from
http://www.ultranet.com/\~{}jkimball/.}
\end{figure}

\subsection{Relevance to biology}
Clearly, random genome realizations ar far from being a realistic model of real biological
genetic systems. However, it can be shown that even this extreme
oversimplification has some relevance for biology.  In Figure ~\ref{fig:numgenes}, the
predicted number of genes in a genome, $N=(1/4)^{l_p} \cdot S$, is plotted as a
function of genome size for $l_p=5$. Observed data
from 50 prokaryotic organisms that have been completely sequenced are also
shown. The correspondence between model and data is excellent for
this range of $S$ and shows that a combinatorial method for
determining the number of genes in a genome is appropriate. For larger
$S$, as typically found in eukaryotic
organisms, $l_{p}=7$ is reasonable (not shown), but little observed data exists.
On the other hand, statistical distributions of regulatory inputs and outputs
do not match biological data particularly well; here, more
realistic statistics can be obtained by constructing artificial genomes
from duplication and divergence events \cite{Leier2007}. However, even in these
models the question how selection pressure on the {\em phenotype},
as encoded by network dynamics, may influence genome organization,
remains unanswered. This type of question shall be addressed in the remaining part
of this paper.

\section{Stabilizing selection for a phenotype - an evolutionary scenario}
Though evolved by the random processes of genetic drift and selection pressure from changing environments, real genetic
systems are far from being random. Complex organization in genome structure is often connected to the  highly non-linear
nature of the genotype-phenotype map, which includes an intermediate layer of complex regulatory processes
controlling cell machinery (unicellular organisms) or highly structured developmental processes (multicellular
organisms). The multilevel-structure of the involved evolutionary processes is sketched schematically in Fig. 4.:
the genome, i.e. the DNA sequence, codes not only for structural proteins, but also for a complex gene
regulatory network (GRN). The dynamics of this GRN regulates the development of the phenotype. The environment
influences development twofold, first by perturbations of the developmental process (noise), second by
selection pressure for viable phenotypes. Organisms reproduce by duplication of their genome, which is
an imperfect process frequently leading to errors (mutations).
Typically, models of evolutionary adaptation focus either on sequence evolution or network structure alone,
and hence imply a huge loss of information as compared to the true multi-level and multi-scale evolutionary
dynamics.
\begin{figure}[th]
\centerline{\psfig{file=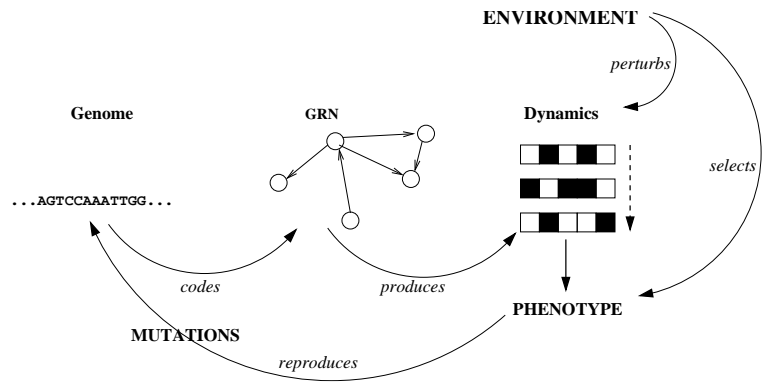,width=10cm}}
\vspace*{8pt}
\caption{Multilevel structure of the evolving genotype-phenotype map: besides coding for structural proteins,
genomes also encode their own regulation by complex gene regulatory networks (GRN). GRN control
the temporal and spatial dynamics that leads to "production" (development) of the organism (phenotype).
The phenotype is reproduced by genome duplication, involving mutations. The environment can perturb developmental
dynamics, as well as it selects for viable/adaptive phenotypes.}
\end{figure}
Artifical genomes could be an important step towards models that integrate these levels, and hence may lead
to predictions on the effects of adaptive processes on sequence- and network evolution, and how these are related
to each other. 

\subsection{Definition of the evolutionary algorithm}
In this section, we briefly explore an example of an evolutionary scenario based on an artificial genome,
motivated by the observation that development is highly canalized, i.e. buffered against both intrinsic and environmental
noise, and mutations \cite{Siegal2002}. A number of studies has demonstrated that stabilizing selection for particular phenotypes
leads to emergence of this high {\em robustness}, strongly facilitated by the high amount of neutrality contained
in the fitness landscapes of complex regulatory networks \cite{Fernandez2007}. Let us now define an evolutionary algorithm of stabilizing selection
in a strongly fluctuating environment, based on an artificial genome.
We start by generating an initial population of randomly assembled genomes; the number of bases $S$
is constrained such that each string contains exactly 64 genes. In all simulations discussed in the following,
a promoter length $l_p = 4$ and a gene length $l_g =6$ are applied. Next, different limit cycles of the
associated RTNs are identified by running network dynamics, as defined in section 2.1.1, from $10^4$ different
random initial state configurations. This process is stopped when a RTN is identified which has a fixed point
$S_f$ (a limit cycle  of length one). In addition, we require that there can be identified at least 4
additional attractors, motivated from phenotype diversity frequently observed in many species \cite{Kussell2005}.
Adaptation  to unpredictable environments is often handled by stochastic switching between phenotypes and can
lead to stabilization of even very small subpopulations of phenotypes that differ from the
population majority \cite{Kussell2005}, which we model by the
requirement that the relative weight of the basin of attraction
leading to $S_f$ should be small (less than $40\%$ of the tested configurations). 
$S_f$ is the phenotype
we want to stabilize, and the digit string $G_f$, that codes for its regulatory network, is the genotype we evolve.

We now apply stabilizing selection for $S_f$ as follows:
\begin{enumerate}
\item Create a mutant $\bar{G}_f$ by random single base mutations, occurring with a probability $p_m = 0.001$ per base.
\item Run RTN dynamics from a random initial state, until an attractor is reached, otherwise stop after 200
iterations.
\item If dynamics has converged to $S_f$, keep $\bar{G}_f$, otherwise keep $G_f$.
\item For the next generation, iterate from (1).
\end{enumerate}
We note that we disregard mutations of promoter sites, as well as mutation leading to new "genes", to avoid
complications resulting from a varying genome size.
Notice that, in step (2), we test only one initial configuration, corresponding to the fact that
biological organisms are tested only against the environment they face at the {\em current} generation.
Robustness against fluctuations, i.e. the capacity to stabilize the phenotype under diverse perturbations
of development by unpredictable environmental influences as well as internal noise,
is measured by running RTN dynamics for $G_f$ ($\bar{G}_f$) for a larger set $Z$ of initial configurations (e.g. $10^4$ random initial states). This variation in initial states simulates the fact that neither all a organisms in one generation meet
a homogeneous environment, nor environments are constant over the course of generations.
Then 
\begin{equation}
R_f(t) := \frac{Z_f(t)}{Z}
\end{equation}
defines the robustness against fluctuations, where $Z_f(t)$ is the fraction of initial states that lead to $S_f$
at generation $t$.
A second measure of robustness is associated to the capacitance to buffer the system against disadvantageous
mutations (mutational robustness $R_m$, \cite{Ciliberti2007}). At each generation we measure
the number of accepted mutants $P_a$ in the previous $P$ generations, and define
\begin{equation}
R_m(t) := \frac{P_a(t)}{P}.
\end{equation}

If $P_a$, and hence $R_m$ increases with $t$, this indicates restructuring of the genome such that
the probability of neutral or advantageous mutations with respect to $S_f$ has increased.

\begin{figure}[th]
\centerline{\psfig{file=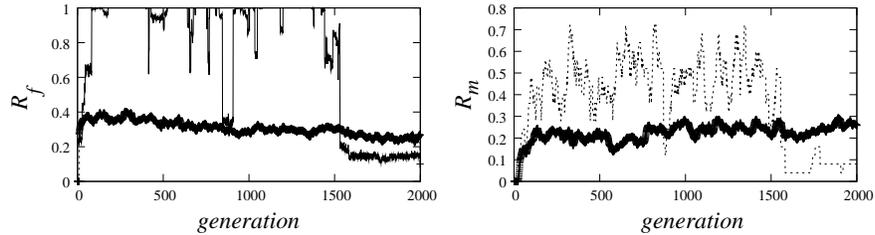,width=11.5cm}}
\vspace*{8pt}
\caption{Time course of evolutionary dynamics. Left: Evolution of the robustness $R_f$ against fluctuations
in initial conditions, example of a particular run (thin-lined curve) and ensemble-average over 67 different
evolutionary runs (thick-lined curve).  Right: Evolution of the mutational robustness $R_m$, the dashed curve
represents an example of an evolutionary run, the thick-lined curve the ensemble average. \label{evodyn_timefig}}
\end{figure}

\begin{figure}[th]
\centerline{\psfig{file=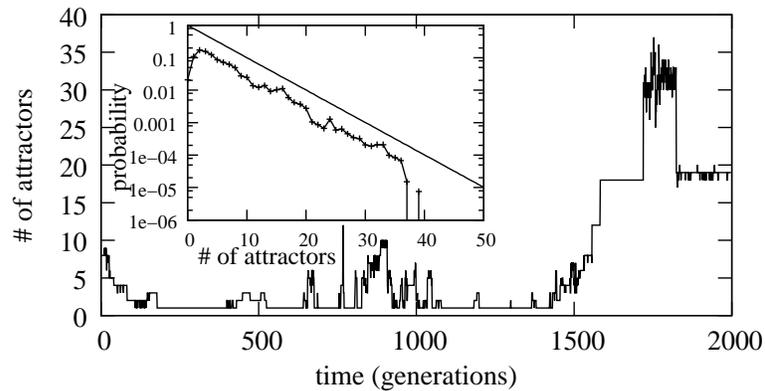,width=10cm}}
\vspace*{8pt}
\caption{Number of different dynamical attractors, as identified by sampling
dynamics from $10^4$ random initial conditions in each generation, in a typical evolutionary run.
Inset: statistical distribution of the number of indentified attractors, sampled from
67 evolutionary runs. The distribution exhibits an exponential decay (straight line
shown for eye guidance). Notice that in about $2\%$ of the cases, no attractors
was identified due to the imposed length constraint on dynamical trajectories.\label{attrnumevo_fig}}
\end{figure}

\begin{figure}[th]
\centerline{\psfig{file=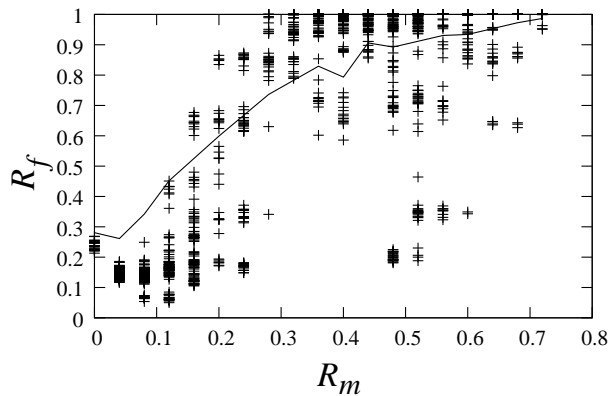,width=8cm}}
\vspace*{8pt}
\caption{Robustness $R_f$ against fluctuating dynamical initial conditions as a function of the
robustness $R_m$ against point mutations of the genome sequence. Crosses show example results of a single
evolutionary run, the lined curve is the ensemble average over 67 different runs. $R_f$  and $R_m$ exhibit a
clear positive correlation. \label{rrcorrfig}}
\end{figure}

\subsection{Results}
Next, let us summarize the results obtained from evolutionary runs, starting from
different random genome realizations with parameters as outlined in the previous
paragraph.
\subsubsection{Evolution of robustness}
Figure \ref{evodyn_timefig} shows both quantities in a typical evolutionary run, and ensemble averages
obtained from 67 evolutionary runs starting from different random genome realizations.
Both $R_f$ and $R_m$ increase rapidly,
however, exhibiting considerable fluctuations. In particular, $R_f$ exhibits an interesting
intermittent dynamics reminiscent of a {\em punctuated equilibrium} \cite{Bornholdt1998}, indicating
metastability of the evolutionary dynamics. 
In fact, in most evolutionary runs we studied $R_f$ and
$R_m$ could be stabilized only over a finite number of generations, as indicated in Fig. 
by the sharp decrease of both quantities around $t = 1500$. 
The metastability is also visible in the ensemble average of $R_f$ (Fig.  \ref{evodyn_timefig},
left panel), which, after an initial sharp increase up to $R_f \approx 0.4$ shows a slight decline
over the following generations. Another measure that can be applied
to characterize the evolution of network dynamics is the number of different attractors (limit cycles)
that are identified by the evolutionary algorithm in successive generations. Ideally,
when stabilizing selection always succeeds, only the fixed point attractor corresponding
to the phenotype $S_f$ should be present. Figure \ref{attrnumevo_fig} shows
that indeed most of the time the number of attractors is very small, however, there are intermittent
increases (bursts). The inset of Fig. \ref{attrnumevo_fig} shows that the statistical
distribution of this quantity, as obtained from multiple evolutionary runs,
exhibits an exponential decay. 
As it will be discussed later on, the "punctuated equilibrium" of evolutionary dynamics is both
related to the selection criterion we chose, and to the mutation dynamics of the artificial genome,
which is considerably different from single-link rewiring, as applied in most comparable
"network only" studies. 

$R_m$ is a measure of the probability that mutations are advantageous or neutral.
In particular, neutrality of mutations strongly facilitates to find better phenotypes,
since it allows evolution to explore a large number of different system
configurations potentially leading to better phenotypes. 
When neutrality is a driving force of evolutionary dynamics,
we expect that robustness against deleterious mutations and fitness of evolved
phenotypes, i.e. $R_m$ and $R_f$, are correlated. 
As becomes evident from Fig. \ref{rrcorrfig}, 
$R_f$ and $R_m$ are indeed positively
correlated, similar to the results reported e.g. in \cite{Ciliberti2007}.  

The artificial genome now allows us to further investigate the effects of this evolutionary dynamics on
both network and sequence structure.

\begin{figure}[th]
\centerline{\psfig{file=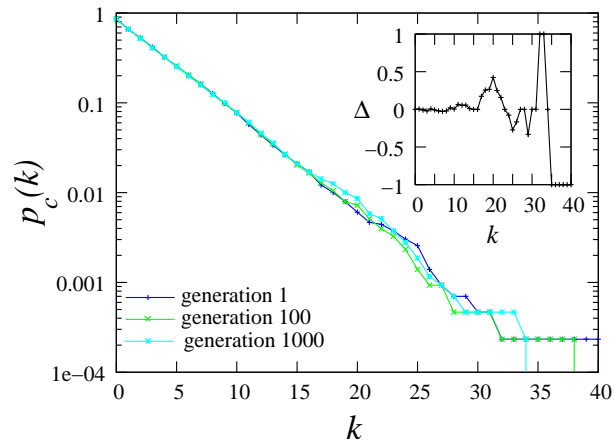,width=8cm}}
\vspace*{8pt}
\caption{Cumulative degree distributions $p_c(k)$ of incoming regulatory links (indegree distribution)
for generation $1$, $500$ and $1000$, averaged
over $100$ evolutionary runs. Inset: Relative difference $\Delta:= (p_c(t=1000)-p_c(t=1))/p_c(t=1)$ between the distributions
at generation 1 and 1000. Notice the increase around $k =18$, followed by a decrease for $k>20$ (details
are discussed in the text). \label{evo_indgree_fig}}
\end{figure}

\begin{figure}[th]
\centerline{\psfig{file=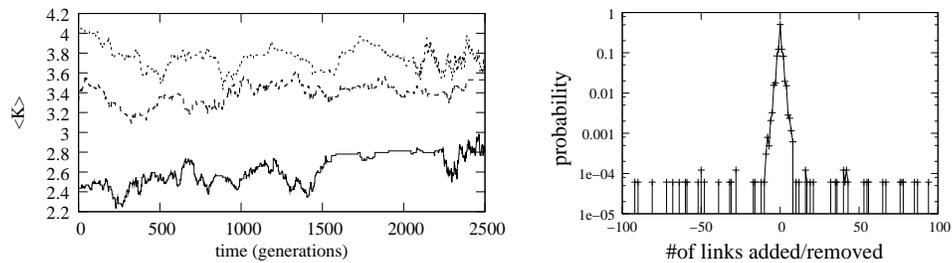,width=12.5cm}}
\vspace*{8pt}
\caption{Left panel: Evolution of the average network connectivity in three different runs. Right panel:
statistical distribution of the number of links added/removed by point mutations
in successive generations of evolving networks. Statistics was taken over all accepted mutants,
averaged over 67 different evolutionary runs. About $50\%$ of mutations are structurally
neutral (no link added or removed), the rest shows a broad spectrum of rewiring effects
at the network level.\label{averkevo_fig}}
\end{figure}

\begin{figure}[th]
\centerline{\psfig{file=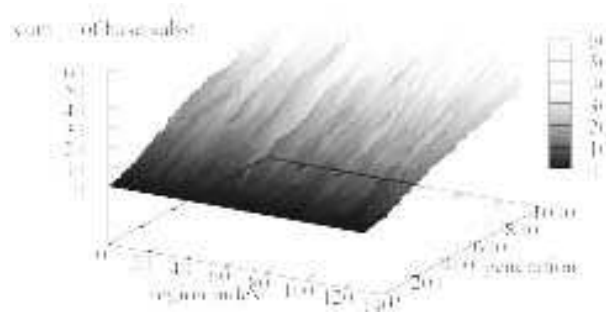,width=8cm}}
\vspace*{8pt}
\caption{Cumulative number of base exchanges during evolution for different positions on the genome,
averaged over regions containing 100 bases each, during the course of evolution from generation 1 to 1000
in a particular evolutionary run.
The brightness in grayscale indicates the number of bases exchanges. Increasing ruggedness of the
surface points at divergent evolution of genomic regions accumulating base changes with rates that differ
by orders of magnitude.}
\end{figure}

\begin{figure}[th]
\centerline{\psfig{file=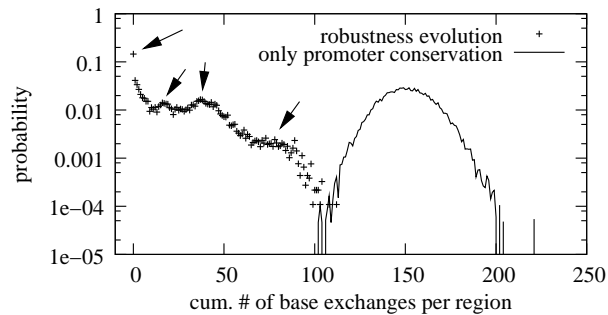,width=8cm}}
\vspace*{8pt}
\caption{Statistical distribution of the cumulative number of base exchanges per region at generation
$2000$, averaged over $100$ different evolutionary runs. Left curve (+): evolution with the dynamical
(robustness) constraint as described in the text. The distribution has a complex multimodal structure
with maxima/plateaus around $0$, $40$ and $80$ (arrows). The lined curve on the right shows the same distribution
for dynamically unconstrained evolution (keeping only promoter sequences fixed).}
\end{figure}

\subsubsection{Evolution of network- and sequence structure}

Let us first look on the evolution of network structure under the imposed {\em dynamical} robustness constraint.
We performed $100$ evolutionary runs with different initial assignments of $G_f$ and $S_f$; each simulation
was observed over 2000 generations, and regulatory networks evolved after 2000 generations were
compared to the initial networks. With regard to average network connectivity and outdegree distributions,
no substantial reorganization was found. However, moderate reorganization is found in the distribution
of regulatory {\em inputs}, which is shown in Fig. \ref{evo_indgree_fig} (for smoothing of data,
cumulative distributions $p_c(k) :=$ probability to observe a node with indegree $\ge k$ are shown,
besides averaging over 100 different evolutionary runs). Typically, an increase of probability for
intermediate values of $k$ is found, while probability for larger $k$ is reduced. This can be clearly
appreciated by investigation of the difference between both distributions (inset of Fig. \ref{evo_indgree_fig}).
However, the overall shape of the distribution does not change significantly and still stays close to an exponential,
presumably due to the small network sizes and limited number of generations observed.
Figure \ref{averkevo_fig} (left panel) shows the time evolution of the average network connectivity $\bar{K}$ for three
different evolutionary runs. $\bar{K}$ shows considerable variance both with regard to different
initial random genome realizations evolution starts from, as also with regard to fluctuations
during evolution. In particular, the effects of point mutations at the sequence level on
network wiring are strongly non-linear. To show this, we measured the statistical distribution
of the number of regulatory links that were deleted or added in successive generations of accepted
mutants, averaged
over 100 different evolutionary runs (Fig. \ref{averkevo_fig}, right panel). In about $50 \%$
of the cases, mutations did not affect network wiring; in the remaining cases, most often
only one or a few links were affected (see the sharp peak around zero), however, there
are also cases were a large number of links is added or removed simultaneously.
This result is in contrast to many other studies of network evolution, which implicitly
assume small, stepwise local changes in network wiring and hence only small
moves along neutral paths of the fitness landscape. In our model,
this is still the most frequent case, however, in some instances also larger jumps between
different peaks of the fitness landscape naturally emerge through mutations in the
sequence-based encoding of network structure that affect a large number of regulatory
interactions. 

Last, let us investigate how evolution proceeds at the most basic level of the
system, i.e. the digit sequence of the artificial genome.
Figure 10 shows the number of base exchanges during evolution for different positions on the genome.
At each generation, the cumulative number of base substitutions in successive slices of 100 digits on the genome string,
identified by a unique region index,
during all previous generations was monitored. Increasing ruggedness of the
surface points at divergent evolution of genomic regions accumulating base changes with at very different rates,
giving evidence that there co-exist highly conserved and "adaptive" regions. We hypothesize that the former
encode the invariant "core" of the regulatory network needed to produce the phenotype, while the latter
contain neutral mutations, or support buffering against fluctuations. In Fig. 11, the statistical distributions
for cumulative number of base substitutions in evolved genomes are compared to the control experiment
without robustness constraint, only requiring preservation of promoter- and gene sequences.
While in the control experiment, a simple, symmetric binomial distribution is found, evolved networks
exhibit a strongly asymmetric, multi-modal distribution with at least three identifiable maxima/plateaus
(indicated by arrows). This demonstrates that the selective constraint on regulatory dynamics indeed 
strongly influences the evolutionary patterns found in the genome at sequence level.

\section{Discussion and Conclusions}
We studied statistical properties of the artificial genome model proposed by Reil \cite{Reil1999}
both on the level of base sequences and regulatory networks generated with this model,
and the evolution of developmental canalization (robustness against noise
 in initial conditions of regulatory
dynamics and against single base mutations). We find that
random realizations of the artificial genome exhibit pronounced differences
between the statistical distributions of regulatory inputs and outputs,
and a scaling in the number of genes (as a function of the number of DNA bases)
compatible with corresponding data of prokaryotic organisms for choices
of model parameters as typically applied in our simulations.
The simulation of evolutionary dynamics yields a number of surprising results.
First, we observe that robustness is an evolvable property, and in particular
that robustness against deleterious mutations and robustness against noise
are correlated, similar to results of other studies \cite{Ciliberti2007}.
However, while in most "network only" studies (without a sequence based
description) only small adaptive changes (rewirings) are considered,
we find emergence of highly non-linear effects between sequence point mutations
and network wiring (as predicted in \cite{Watson2004a}
for random genome realizations), including a large number of structurally neutral mutations,
and mutations that lead to simultaneous rewiring of multiple connections.
This means that, while stepwise evolution along neutral paths of 
the  fitness landscape with regard to phenotypic effects of mutations
\cite{Bornholdt1998,Ciliberti2007,Fernandez2007} is still the main driving
mechanism, also larger jumps between different peaks of 
the  fitness landscape are possible. Interestingly, we found
evidence that this increased non-linearity in the genotype-phenotype
map and the resulting fitness landscape
tends to weaken the effectiveness of stabilizing selection in the long run,
and the degree of evolved robustness exhibits considerable
fluctuations during evolutionary runs.
Concerning network structure, we found only moderate reorganization
of the statistical distributions of input- and output numbers per
node during evolution. In contrast,
evolution leaves clearly visible signs at sequence level
with a pronounced pattern of strongly conserved regions, and other parts of the genome evolving
in a much less constrained way.

To conclude, the results of our study indicate that artificial
genomes represent an interesting step towards more realistic
models for the evolution of gene regulatory networks (GRN),
by taking into account the indirect evolution of GRN structure
through mutation of regulatory sequences, which cannot be accounted
for in "network only" models. 
Clearly, the results of the current
study are limited in scope; in future extensions of the model,
we will in particular address variations in the number of
genes (e.g. resulting from sequence duplications), and more
realistic models for the binding of transcription factors
to regulatory binding sites.

\bibliography{agenome_bib1_140408}{}
\bibliographystyle{plain}

\end{document}